\documentstyle[11pt,paspconf]{article}

\begin{document}

\title{Gravitational lensing in galaxy redshift surveys}

\author{Daniel J.\ Mortlock \& Rachel L.\ Webster}
\affil{School of Physics, University of Melbourne, 
	Parkville, Vic.\ 3052, Australia}


\section{Introduction}

Gravitationally-lensed quasars are both valuable (individually
and statistically) and rare.
They are mostly discovered through time-consuming 
re-imaging of known quasars, but can also be found 
spectroscopically in galaxy redshift surveys.
Q~2237+0305 (Huchra et al.\ 1985) was found in this manner,
but remains the only such discovery.
Kochanek (1992) predicted that $\sim 1$ in $10^5$ redshifts would 
yield a lens although this is somewhat pessimistic, as 
described below.
It is also important that lenses discovered in galaxy surveys
tend to have bright, low redshift lens galaxies.
Such lenses are particularly useful, as shown by 
the successful measurement of the 
source-size and transverse velocity in the 
Q~2237+0305 system (e.g., Wambsganss et al.\ 1990; Wyithe et al.\ 1999).

\section{Calculation}

The calculation,
detailed in Mortlock \& Webster (1999),
of the expected number of lenses, $N_{\rm lens}$,
is based on the formalism developed in
Kochanek (1992). 
The lensing probability of a single galaxy is integrated over 
the survey population,
subject to a number of selection effects.
A lens of total magnitude $m$,
in which the quasar images have magnitude $m_{\rm q}$ 
and the galaxy has magnitude $m_{\rm g}$, is detectable 
in a galaxy redshift
survey if:
the quasar-galaxy composite image is brighter than the survey
limit (i.e., $m \leq m_{\rm lim}$);
the quasar images are bright enough to be detected 
in the composite spectrum (which can be quantified by 
requiring that
$m_{\rm q} - m_{\rm g} \leq \Delta m_{\rm qg} \simeq 2$; 
Kochanek 1992); 
and the galaxy is bright enough for the composite object to be 
classified as non-stellar, and hence enter the redshift survey
at all (which can be characterised by the condition that
$m_{\rm g} - m_{\rm q} \leq \Delta m_{\rm gq} \simeq 0$).
The inclusion of the quasar light in the calculation of 
$m$ is critical to this calculation, as the 
sensitivity to lenses is increased by $\sim 1$ mag,
and the number of lenses is increased by up to an order of magnitude,
relative to the results of Kochanek (1992). 

\section{Results}

\begin{figure}
\includegraphics{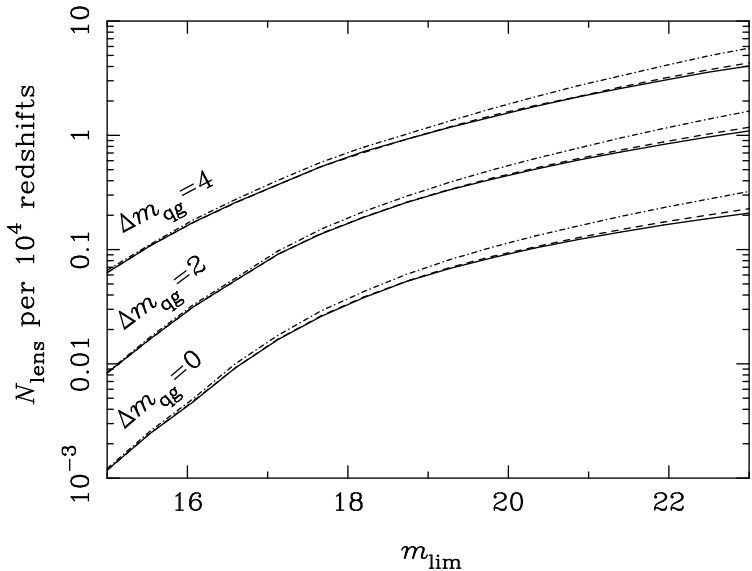}
\vspace{7.8cm}
\caption{The expected number of lenses,
$N_{\rm lens}$, per $10^4$ measured redshifts
in a galaxy redshift survey of magnitude limit $m_{\rm lim}$. The three
sets of lines are for different values of the ``spectral prominence''
parameter, $\Delta m_{\rm qg}$, as indicated.
In each case results are shown 
for three cosmological models: 
Einstein-de Sitter (solid lines); 
empty (dashed lines); 
and empty and flat (dot-dashed lines).}
\label{figure:n_lens}
\end{figure}

Figure~\ref{figure:n_lens} shows the number of lenses expected
per $10^4$ redshifts as a function of survey depth.
Several values of $\Delta m_{\rm qg}$ are shown,
as this represents the largest single uncertainty in the calculation.
Note that, at least at the depths of the current redshift surveys,
there is very little dependence on the cosmological model, 
which stands in marked contrast to the usual
quasar lensing probability.
These results imply that the a posteriori likelihood of a lens 
being discovered in the Center for Astrophysics (CfA) survey 
is $\sim 0.03$, which is much greater than 
the previous estimate of $\sim 0.003$ (Kochanek 1992).

The new generation of surveys -- in particular the 
Sloan Digital Sky Survey (SDSS)
and the 2 degree Field (2dF) redshift survey 
--
will dwarf the CfA survey, and should 
result in the discovery of many new lenses. 
The SDSS will obtain 
$\sim 10^6$ high quality galaxy spectra 
(i.e., $\Delta m_{\rm qg} \ga 4$)
to $m_{\rm lim} \simeq 19$, which implies $N_{\rm lens} \simeq 100$.
The 2dF redshift survey
will contain a quarter the number of galaxies, but has 
$m_{\rm lim} = 19.5$, and should yield $\sim 10$ lenses.
Most of these lenses will have deflector redshifts of 
$\la 0.2$, and 
in several cases the observer-deflector distance should be 
comparable to that in the Q~2237+0305 system.

\end{document}